\documentclass{article}

\usepackage{PRIMEarxiv}

\usepackage[utf8]{inputenc} % allow utf-8 input
\usepackage[T1]{fontenc}    % use 8-bit T1 fonts
\usepackage{hyperref}       % hyperlinks
\usepackage{url}            % simple URL typesetting
\usepackage{booktabs}       % professional-quality tables
\usepackage{amsfonts}       % blackboard math symbols
\usepackage{nicefrac}       % compact symbols for 1/2, etc.
\usepackage{microtype}      % microtypography
\usepackage{lipsum}
\usepackage{fancyhdr}       % header
\usepackage{graphicx}       % graphics
\graphicspath{{media/}}     % organize your images and other figures under media/ folder
\usepackage{csquotes}
\usepackage{float}
\usepackage{threeparttablex}

\title{Foreign participation in federal biddings: a quantitative approach using the Procurement Panel
\thanks{{\textbf{Original article in Portuguese. Cite as}}: 
\textit{Ferreira, Carlos. Participação estrangeira em licitações federais: uma abordagem quantitativa utilizando Painel de Compras. Revista Do Serviço Público, 72(4), 779-802. DOI: 10.21874/rsp.v72.i4.4628}} 
}

\author{
  Carlos Ferreira \\
  Faculty of Engineering \\
  University of Porto \\
  Porto\\
  \texttt{ferreira.carlos@fe.up.pt} \\
}

\begin{document}
\maketitle

\begin{abstract}
The bidding is the Public Administration's administrative process and other designated persons by law to select the best proposal, through objective and impersonal criteria, for contracting services and purchasing goods. In times of globalization, it is common for companies seeking to expand their business by participating in biddings. Brazilian legislation allows the participation of foreign suppliers in bids held in the country. Through a quantitative approach, this article discusses the weight of foreign suppliers' involvement in federal bidding between 2011 and 2018. To this end, a literature review was conducted on public procurement and international biddings. Besides, an extensive data search was achieved through the Federal Government Procurement Panel. The results showed that between 2011 and 2018, more than R\$ 422.6 billion was confirmed in public procurement processes, and of this total, about R\$ 28.9 billion was confirmed to foreign suppliers. The Ministry of Health accounted for approximately 88.67\% of these confirmations. The Invitation, Competition and International Competition modalities accounted for 0.83\% of the amounts confirmed to foreign suppliers. Impossible Bidding, Waived Bidding, and Reverse Auction modalities accounted for 99.17\% of the confirmed quantities to foreign suppliers. Based on the discussion of the results and the limitations found, some directions for further studies and measures to increase public resources expenditures' effectiveness and efficiency are suggested.
\end{abstract}

% keywords can be removed

\section{Introduction}
Bidding can be understood as an administrative procedure used by the persons appointed by law to select the most advantageous proposal for public administration through objective and impersonal criteria (Oliveira, 2015). Brazilian legislation provides for the holding of international bidding, which allows foreign companies to accede to the administration's call for the supply of goods or services (Brazil, 1993). In addition, there are understandings that the possibility of participation of foreign bidders is not restricted to international bids. In theory, domestic bids could also admit the involvement of interested parties based in other countries (Schwind, 2013).

The legal system has no formal definition for the term "international bidding" (Schwind, 2013). However, it can be said that, in international bidding, companies based in other countries are admitted to participate, even if this occurs in Brazil (Pereira, 2013).

The competent authority, on certain occasions, can opt for national or international bidding according to its discretion. In other events, however, it may not use its discretion, so it must opt for international or national bidding in a mandatory way (Pestana, 2013). 

In the words of Professor Marcio Pestana:

\begin{displayquote}
"The very need of the public administration, ontologically considered, decisively influences the choice of the type of bidding. The urgency of contracting a specific service or acquiring a particular product certain product that, here in Brazil, is recognized as being well provided or is presented with attractive quality and price practically limits the domain of administrative merit of the authority since it will have no alternative but to conduct the national bidding. At the other extreme, however, if the demand falls on products inexistent in the country or on services that can only be provided quality, reasonable price and delivery time compatible with the needs of administration abroad, then such conditions will also limit the agent's margin of discretion, requiring that the bidding be international type." (Pestana, 2013, p. 371-372)
\end{displayquote}

Contracts of an international nature may involve three distinct hypotheses which require caution from the manager (Pereira, 2013). The first hypothesis deals with international bids financed with domestic resources, i.e., resources coming from the Brazilian government (Schwind, 2013).

The second is provided in paragraph 5 of the art. 42 of Federal Law N.º 8,666/1993 (Bidding Law) refers to the contracting of works, services or goods with resources from international bodies of which Brazil is part (Pereira, 2013). These are international bidding financed with resources, for example, the Inter-American Development Bank - IDB and the International Bank for Reconstruction and Development - IBRD or World Bank (Schwind, 2013).

The last hypothesis deals with the Waiver of Bidding provided in art. 24, section XIV, of the Bidding Law, which refers to goods and services acquired through an international agreement when the conditions are favourable to the government. The bidding is waived, and the contracting will be made with the company linked to the international agreement (Pereira, 2013).

In this context, international bidding is increasingly gaining a place in public administration, with suppliers from several countries selling their products to the Brazilian State, through competition among them (foreigners) and the Brazilian bidders (Carvalho, 2014).

However, in general, it has been a theme little deepened by the doctrine, but, because of globalization and the insertion of Brazil in the international market, it becomes essential the study the bids promoted by the administration that allow the participation of foreign suppliers (Pereira, 2013).

In this sense, the objective of this article is to discuss, through a quantitative approach, the weight of foreign suppliers' participation in the context of federal biddings, between 2011 and 2018, from the use of the tool Procurement Panel, from the Federal Government.

\section{Public Procurement}
\label{sec:headings}

Public procurement is recognised as a strategic instrument in delivering public services. Its prominence as a policy instrument is linked to it is their value since, in countries that are members of the Organisation for Economic Co-operation and Development (OECD), they account for percentages between 4 and 14\% of Gross Domestic Product (GDP) (OECD, 2012).

The purchasing sector is a fundamental part of the achievement of the institutional any kind of organization (Faria et al., 2010) and, in this sense, efficiency is one of the main objectives of the public procurement system of the Federal Government (Inamine; Erdmann; Marchi, 2012).

In hiring, the public sector must comply with the content Bidding Law, among other legal provisions (Silveira et al., 2012). In addition, subsection XXI of art. 37 of the Federal Constitution of 1988 states that works, services, purchases, and sales shall be contracted through a public bidding process except for the cases specified in the legislation. Article 3 of the Bidding Law establishes the purposes of the bid proceeding, which are to ensure the observance of the constitutional principle of equality, the selection of the most advantageous proposal for the administration and the promotion of national sustainable development. The same article further establishes that bids shall be processed and judged in strict conformity with the basic principles of legality, the impersonality of morality, equality, publicity, administrative probity, binding to the summoning instrument, and objective judgment and those that are correlated to them. Article 15 establishes that, in the purchases made, the principle of standardization must be observed whenever possible, and the complete description of the good to be acquired without indicating the brand (Brazil, 1993).

Article 22 of the Bidding Law lists the types of bidding procedures: Competition, Price Taking, Invitation, Contest and Auction. Paragraphs 1 through 5 of art. 22 define each modality of bidding. For Contest and Auction, the object contracted makes a choice. In the case of Competition, Price Taking, and Invitation, the selection shall be made based on the estimated value of the contract, as per art. 23, and presented in Table \ref{tab:table1}.

\begin{table}[H]
\centering
\caption{Bidding modalities and value limits for use (R\$) \tnote{1}}
\label{tab:table1}
\begin{tabular}{cccc}
\hline
\textbf{Object/Modality} & \textbf{Invitation} & \textbf{Price Taking} & \textbf{Competition} \\ \hline
Engineering Works and Services & \begin{tabular}[c]{@{}c@{}}up to\\ 330 000.00\end{tabular} & \begin{tabular}[c]{@{}c@{}}up to\\ 3 300 000.00\end{tabular} & \begin{tabular}[c]{@{}c@{}}more than\\ 3 300 000.00\end{tabular} \\
\begin{tabular}[c]{@{}c@{}}Purchases and Services (except Works and\\ and Engineering Services)\end{tabular} & \begin{tabular}[c]{@{}c@{}}up to\\ 176 000.00\end{tabular} & \begin{tabular}[c]{@{}c@{}}up to\\ 1 430 000.00\end{tabular} & \begin{tabular}[c]{@{}c@{}}more than\\ 1 430 000.00\end{tabular} \\ \hline
\end{tabular}
\begin{tablenotes}
       \item [1] Source: Bidding Law.
 \end{tablenotes}
\end{table}

Concerning international biddings, art. 23, paragraph 3 of the Bidding Law determines that Competition is the method of bidding applicable in international bids. However, the same provision allows the use of others, such as Prices Taking and Invitations. The first must be used when the body or entity has an international list of suppliers. A second is when there is no supplier of the good or service in the country (Brazil, 1993).

In 2002, Federal Law 10,520/2002 (the Reverse Auction Law) established a new type of bidding process called Reverse Auction ("Pregão"). Within the context of changes in public administration, the Reverse Auction procedure emerged as the hope for a more efficient and effective efficient and effective process for bidding procedures. It is faster and more economical since unnecessary steps are eliminated, and the most advantageous proposal is obtained through decreasing bids (Teixeira; Penedo; Almeida, 2012 apud Ribeiro, 2007).

The regulation of the Reverse Auction Law came through Federal Decree N.º 3,555/2000. At first, there was doubt about its permanence since it already regulated Provisional Measure N.º 026/2000, converted into law only in 2002, after several re-editions. However, the suspicion was eliminated hermeneutically since the rules of the decree subsisted when the provisional measure was converted into law, and they remain in the legal system (Bittencourt, 2012).

Subsequently, in 2005, Federal Decree N.º 5,450/2005 regulated the use of the Reverse Auction, in the electronic form, for the acquisition of everyday goods and services when the dispute is held remotely, in public session, through a system that promotes communication over the internet (Brazil, 2005a). Also, in 2005, Federal Decree N.º 5,504/2005 established the requirement to use the Reverse Auction, preferably in the electronic form, for public or private entities to procure common goods and services (Brazil, 2005b).

Recently, Federal Decree N.º 10,024/2019 brought a new regulation for the acquisition of goods and the contracting of standard services, including common engineering services, through the electronic form of Reverse Auction, while it revoked Federal Decree N.º 5,450/2005 and Federal Decree N.º 5,504/2005 (Brazil, 2019a).

With the advent of the Reverse Auction, international biddings are now also conducted through this modality. The discipline of the participation of foreign companies in the Reverse Auction modality is found in art. 16 of Federal Decree N.º 3,555/2000, in art. 15 of the now revoked Federal Decree N.º 5,450/2005 (Bittencourt, 2010; Filho, 2013; Pestana, 2013) and, subsequently, in art. 41 of Federal Decree N.º 10.024/2019.

Article 16 of Federal Decree N.º 3,555/2000 refers to the discipline contained in §paragraph 4 of art. 32 of the Bidding Law. In the words of Professor Marçal Filho:

\begin{displayquote}
"The federal regulation rule overcomes the doubt about the feasibility of foreign companies participating in the Reverse Auction type bidding process. This is the best solution since the legislative silence cannot be interpreted as a prohibition (not even as a lack or absence of authorization) but rather as evidence that the matter was subject to the rules contained in the general legislation." (Filho, 2013, p. 265).
\end{displayquote}

Article 15 of the now revoked Federal Decree N.º 5,450/2005 tried to legitimize international electronic Reverse Auction, in the same way as art. 16 of the regulation confirmed by Federal Decree N.º 3,555/2000 provided for the international face-to-face Reverse Auction, allowing the participation of foreign companies (domiciled in another country) (Bittencourt, 2010). In this sense, Federal Decree N.º 10,024/2019 did not innovate and brought the exact wording of the previous provision, including only, in its sole paragraph, the requirement of sworn translation and apostille, or equivalent consular confirmation of the documents, whether the winner of the bidding is a foreign company.

No obstacle in the legislation prevents the realization of international electronic Reverse Auctions (Bittencourt, 2010 apud Peixoto, 2009). In the words of Professor Márcio Pestana:

\begin{displayquote}
"The Reverse Auction may also be used in international procedures, not only because of the ease of proceeding and simplicity of the object (common good or service) to be contracted. The electronic system with which it is frequently used can stimulate the participation of foreign companies in the bidding process and, consequently, in the subsequent contracting (Pestana, 2013, p. 414).
\end{displayquote}

However, due to the lack of express provision, when holding the Reverse Auction or any other modality, several peculiarities must be observed, in addition to the subsidiary use of the Bidding Law, mainly what is contained in art. 42 (Bittencourt, 2010 apud Peixoto, 2009) and other devices, such as item II, of §1, of art. 3, of the Bidding Law, which provides for equal treatment between Brazilian and foreign companies, except as provided in paragraph 2 of the same article (Brazil, 1993).

Along these lines, another exception, the margin of preference, was included by Law N.º 12,349 of December 15, 2010, resulting from the conversion of Provisional Measure N.º 495/2010, and is provided in subsection I, paragraphs 5 to 13, of art. 3 of the Bidding Law. This margin may be established for domestic services and manufactured products that meet Brazilian technical standards and, according to §8 of art. 3, will be defined by the Federal Executive Branch (Pereira, 2013). Also, according to that article, the margin of preference cannot exceed 25\% (twenty-five per cent) on the price of the products manufactured products and services offered by foreign companies (Brazil, 2010).

\subsection{Foreign funded biddings}

Biddings funded with funds from a foreign source are regulated by international acts or by the rules issued by international organizations whenever these bodies apply their rules as a condition for the granting of financing or donation of funds. It occurs based on paragraph 5 of the art. 42 of the Bidding Law and the international acts to which Brazil is a signatory (Brazil, 1993). These acts may be protocols, conventions, agreements or international treaties, and the rules are usually identified as guidelines. As previously mentioned, some examples of foreign sources of funds are the Inter-American Development Bank (IDB) and the International Bank for Reconstruction and Development (IBRD) or World Bank (Schwind, 2013).

However, despite the rules issued by international funding agencies, doctrine and case law have been frequent in the sense that the fundamental provisions contained in the Federal Constitution and the Bidding Law cannot be set aside. In this sense, Judgment N.º 2,973/2003 - First Chamber of the Federal Court of Accounts - states that even though the contracting derives from IDB loan resources, the fundamental principles provided for in the Federal Constitution, such as competitiveness, economy and legality, must be observed. Similarly, Judgment N.º 239/2007 - First Chamber - states that compliance with World Bank norms and procedures does not exclude the application of the Bidding Law where it does not conflict, always observing the principles of legality and the supremacy of the public interest. Finally, there are several authors, such as, for example, Edmir Neto de Araújo, Toshio Mukai, Jorge Ulisses Jacoby Fernandes and Marçal Justen Filho, who adhere to the same understanding (Pereira, 2013).

\section{Methodology}

\subsection{The discussion on the weight of foreign companies from a quantitative point of view}

This article brings up the subject of international bidding through a quantitative approach and focuses on the discussion of the weight of the participation of foreign suppliers in these acquisitions. To the best of the author's knowledge, such a discussion is unprecedented and justifies its publication. Contributing to this understanding, none of the documents in the literature review approaches the subject from the same perspective.

Some examples are the approach to the competitiveness aspects in international bidding (Castro, 2002), the discussion on the preference margin (Pighini; Gomes, 2013) and the financing in these types of procedures (Ribeiro; Pereira, 2016). In addition, the Brazilian and European scenarios in these types of procedures were compared (Moreira; Guimarães; Torgal, 2015), and the obstacles to opening the Brazilian public market to foreign companies were addressed (Marrara; Campos, 2017). However, none of them focused on the theme of international bidding from the perspective of a quantitative discussion of the weight participation of foreign companies.

\subsection{General Services System - SISG}

The General Services System (SISG) is an integral part of an organic administrative system that encompasses the entire federal public administration to coordinate public logistics activities with a view to greater efficiency. It consists of one among several systems of auxiliary activities of the administration, responsible for the execution of activities of transversal nature (Brazil, 2019g).

It is, in short, the organization, in the form of a system, of the activities of the administration of public buildings and residential properties, material, transportation, administrative communications and documentation, of which the organs and entities of the direct federal, autarchic and foundational administration are part, as provided in paragraph 1, of art. 1, of the Federal Decree N.º 1,094/94 (Brazil, 1994).

\subsection{Federal Government Procurement Panel}

The Government Procurement Panel of the Ministry of Economy, Planning, Development and Management is a tool that presents the central figures on public procurement in one place and aims to offer an overview of public spending and bidding behaviour within the federal administration. It was developed to provide information from all agencies that make up the SISG (Brazil, 2019b).

The panel presents data on biddings, contracts, price registration minutes and prices practised, which allows, in addition to viewing statistical information, to be an essential tool in government transparency, allowing every citizen to create indicators and customized queries as export data in various formats. All this information is available with annualized clippings that can be exported in multiple formats to facilitate analysis (Brazil, 2019b).

\subsection{Using the Federal Government Procurement Panel}

The Procurement Dashboard was used to obtain data on acquisitions made by the Federal Government from 2011 to 2018. The research was conducted in two waves since the system only allows consultation for the current year and the five previous years. The first wave occurred in 2016, when data between 2011 and 2015 were retrieved; the second, in 2019, when data between 2014 and 2018 were retrieved. Data for 2014 and 2015 overlapped as they were recovered at different times. As there was no inconsistency, the data retrieved in the second wave was retained for the sake of chronology.

The data recovery process was simplified. On the home page of the "Purchasing Panel", the "Purchasing Processes" button was clicked to advance to the "Purchasing Processes Panel" interface. The "Do it yourself!" tool was used from this interface, which can be accessed in the upper right corner.

This tool allows the construction of customized reports by defining specific dimensions and metrics. After selecting the period, in the dimension, the parameters year, supplier - CNPJ/CPF, higher body and modality were selected; and, in the metrics, the parameters quantity of purchases and purchase value were selected.

After selecting the parameters, a report preview is shown on the screen. From there, by clicking the right mouse button, it is possible to export the report to Microsoft Excel® format through the option "Send to Excel" in the open menu. After opening the generated file, it is possible to select, already in the spreadsheet, the foreign suppliers, which are designated in the column that refers to the parameter "supplier - CNPJ/CPF" by the denomination of the type "ESTRANGXXXXXXXXX".

\section{Results}

\begin{figure}[h]
  \centering
  \includegraphics[width=0.75\textwidth]{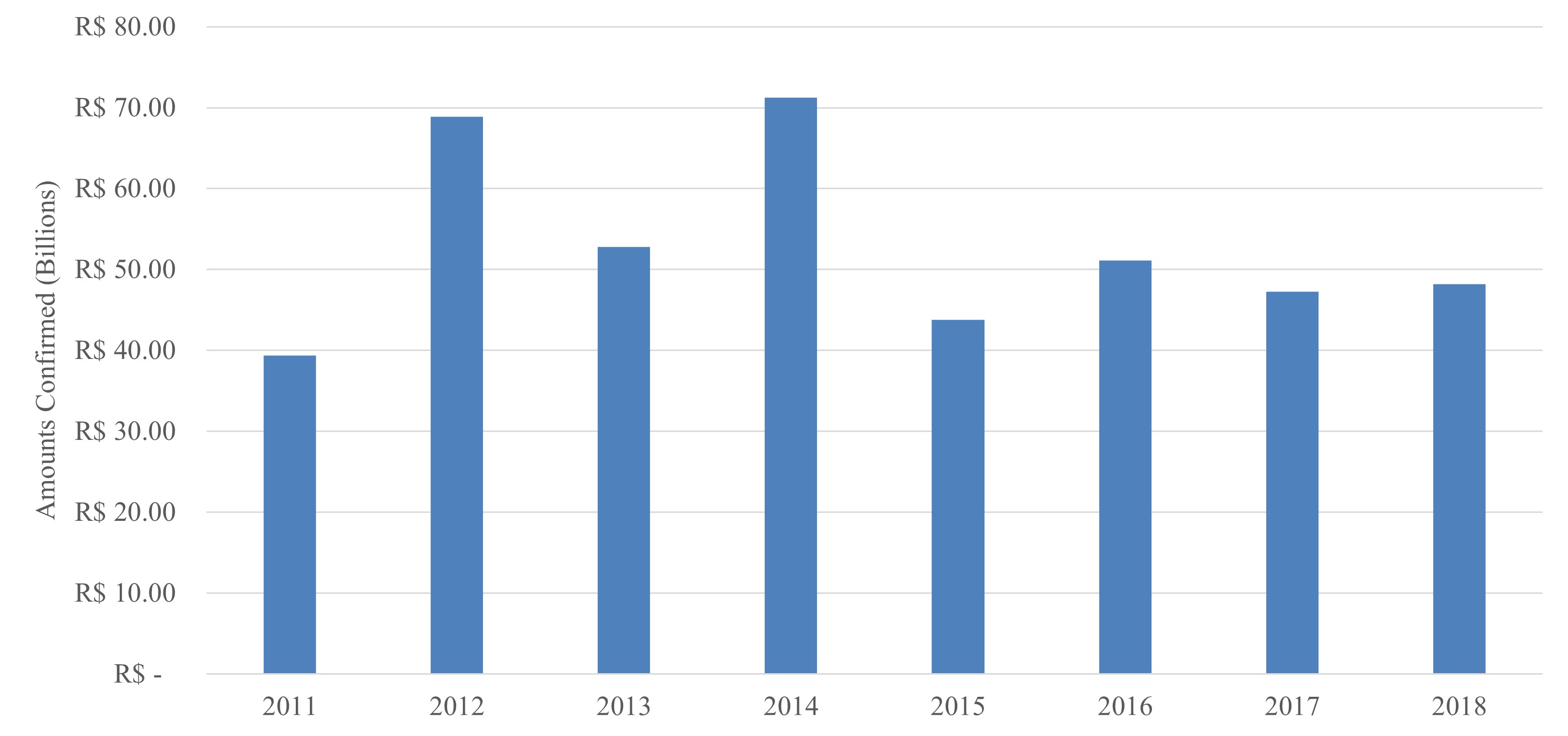}
  \caption{Evolution of confirmed resources in public procurement | Source: Data taken from the Federal Government Procurement Panel} 
  \label{fig:fig1}
\end{figure}

The data obtained from the procurement dashboard were processed, and based on the information generated, the confirmation of more than R\$ 422.6 billion was verified in a total of 1,086,679 public procurement processes within the period considered. The evolution of these resources throughout this period can be seen in Figure \ref{fig:fig1}.

\begin{figure}[h]
  \centering
  \includegraphics[width=0.75\textwidth]{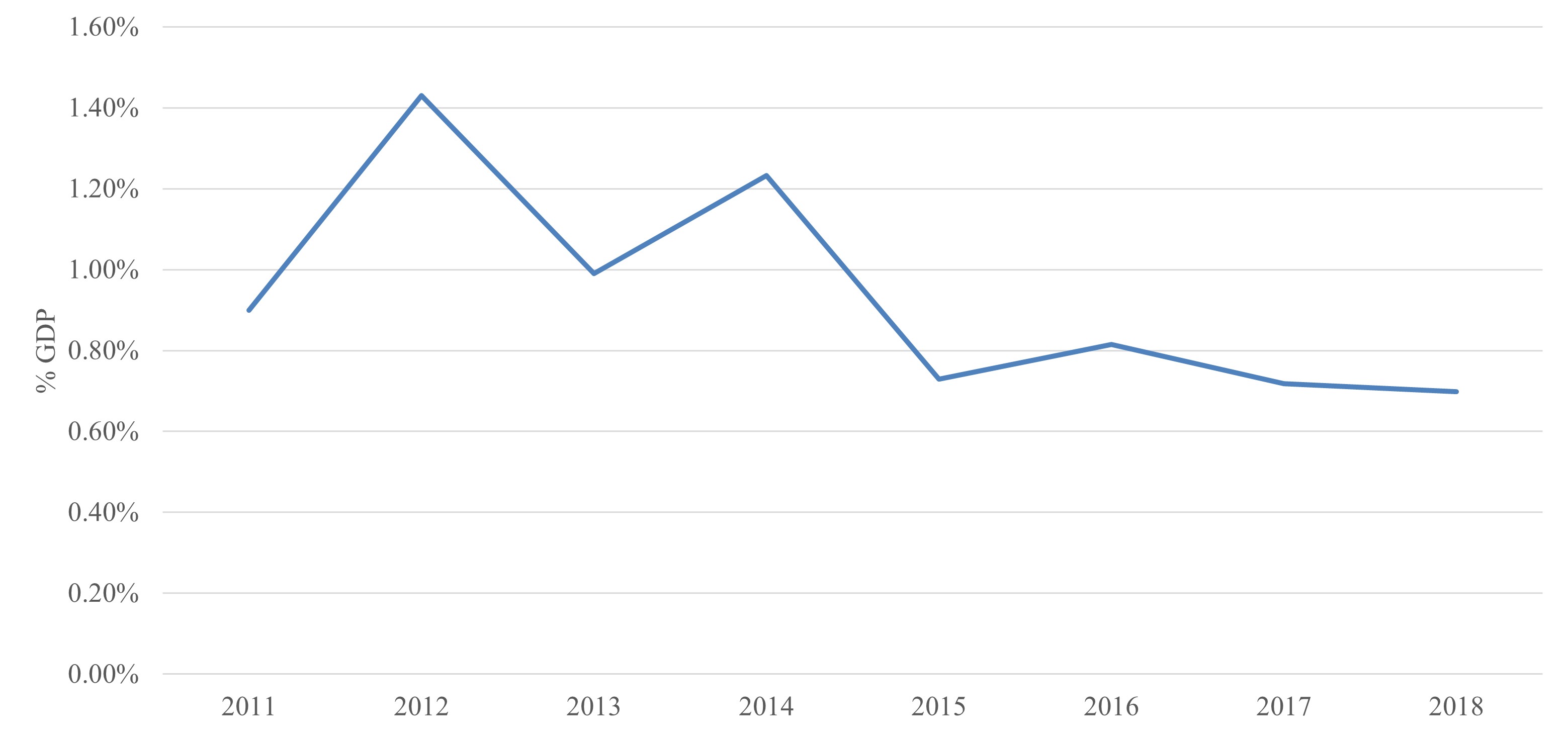}
  \caption{Public procurement as a \% of GDP | Source: Data taken from the Federal Government Procurement Panel}
  \label{fig:fig2}
\end{figure}

The total amount confirmed corresponded to about 0.92\% of the accumulated GDP in the same period, according to data from the Brazilian Institute of Geography and Statistics - IBGE (Brazil, 2019d), with percentages varying between 0.70 and 1.43\% over the years and can be observed in Figure \ref{fig:fig2}.

\begin{figure}[h]
  \centering
  \includegraphics[width=0.75\textwidth]{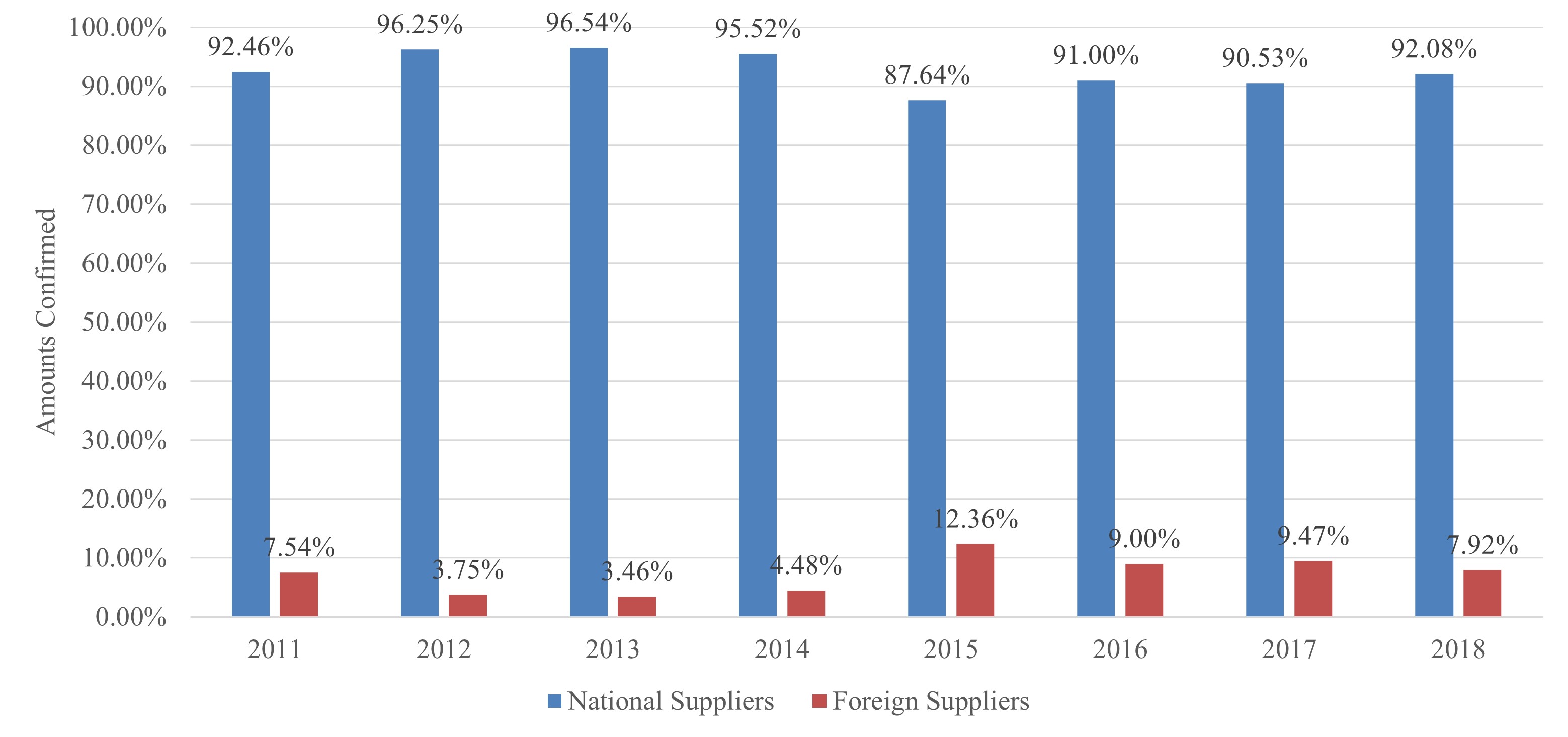}
  \caption{Percentages of resources confirmed to national and foreign suppliers | Source: Data taken from the Federal Government Procurement Panel}
  \label{fig:fig3}
\end{figure}

Approximately R\$2 8.9 billion (6.83\%) was confirmed to foreign suppliers through 22,726 processes. Compared to the total, the annual percentages varied between 3.46\% and 12.36\% over the period. Table \ref{tab:table2} shows the number of procurement processes confirmed to foreign suppliers for modality. Figure \ref{fig:fig3} compares the percentages of resources confirmed to national and foreign suppliers.

\begin{table}[h]
\centering
\caption{Procurement processes confirmed to foreign suppliers \tnote{1}}
\label{tab:table2}
\begin{tabular}{cccccccccc}
\hline
 & 2011 & 2012 & 2013 & 2014 & 2015 & 2016 & 2017 & 2018 & Total \\ \hline
Competition & 2 & 0 & 3 & 0 & 0 & 1 & 0 & 1 & 7 \\
International Competition & 33 & 28 & 42 & 34 & 25 & 23 & 9 & 10 & 204 \\
Inviation & 0 & 0 & 29 & 12 & 0 & 0 & 0 & 0 & 41 \\
Reverse Auction & 58 & 96 & 60 & 55 & 21 & 28 & 18 & 12 & 348 \\
Waiver Bidding & 2,267 & 2,217 & 2,208 & 2,206 & 1,131 & 1,351 & 1,234 & 1,432 & 13,866 \\
Impossible Bidding & 844 & 1,259 & 1,204 & 1,248 & 939 & 752 & 858 & 1,156 & 8,260 \\ \hline
Total & 3,204 & 3,600 & 3,546 & 3,375 & 2,116 & 2,155 & 2,119 & 2,611 & 22,726 \\ \hline
\end{tabular}
\begin{tablenotes}
       \item [1] Source: Data taken from the Federal Government Procurement Panel.
\end{tablenotes}
\end{table}

\begin{figure}[h]
  \centering
  \includegraphics[width=0.75\textwidth]{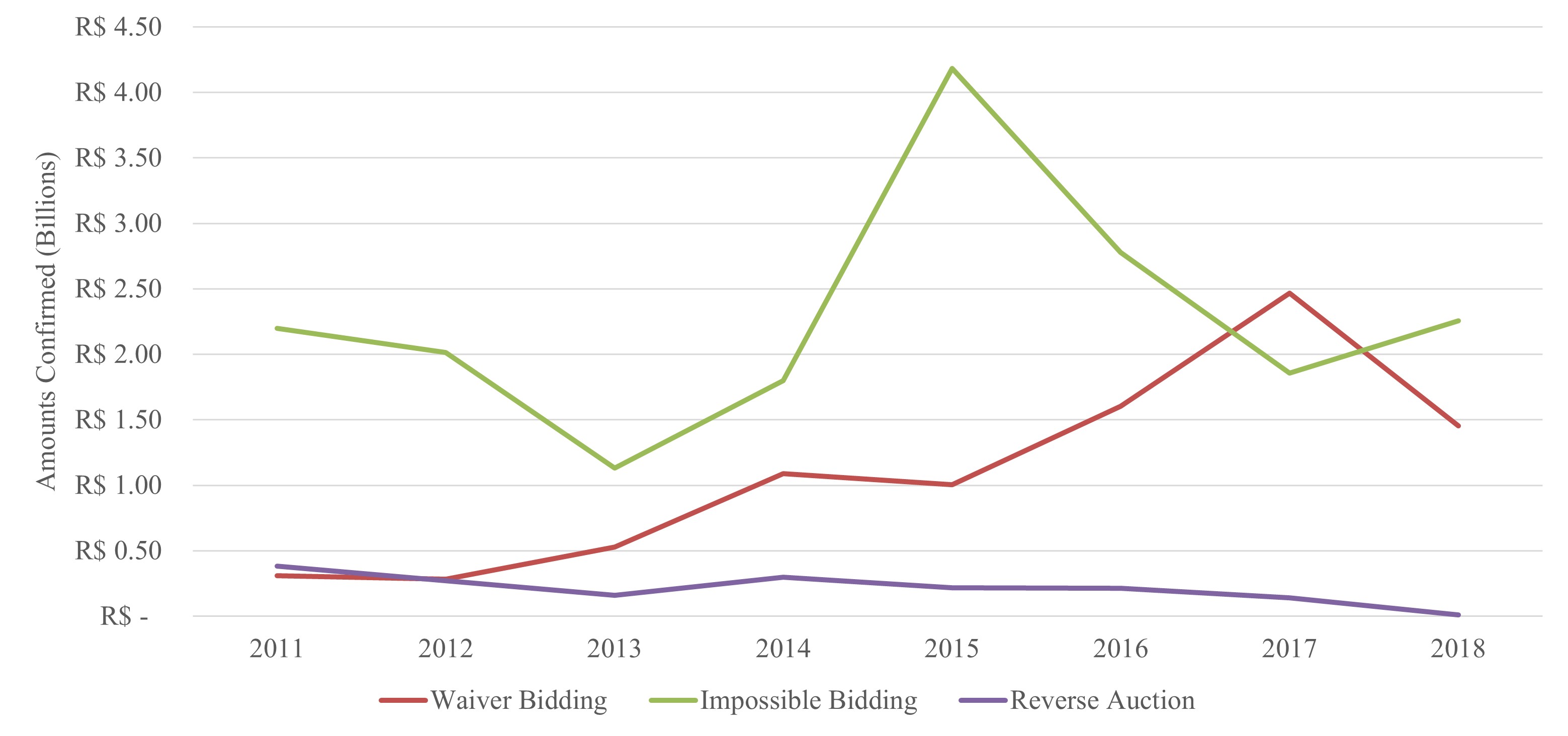}
  \caption{Resources confirmed to foreign suppliers in the Reverse Auction, Waiver and Impossible bidding | Source: Data taken from the Federal Government Procurement Panel}
  \label{fig:fig4}
\end{figure}

The modalities of Invitation, Competition and International Competition accounted for R\$ 239.5 million (0.83\%) of the total resources confirmed for foreign suppliers. The Reverse Auction modality and Waiver and Impossible bidding accounted for about R\$ 28.6 billion (99.17\%) of the total confirmed to foreign suppliers. Figure \ref{fig:fig4} shows the evolution of resources confirmed for foreigners in Reverse Auction, Waiver and Impossible bidding

\begin{figure}[h]
  \centering
  \includegraphics[width=0.75\textwidth]{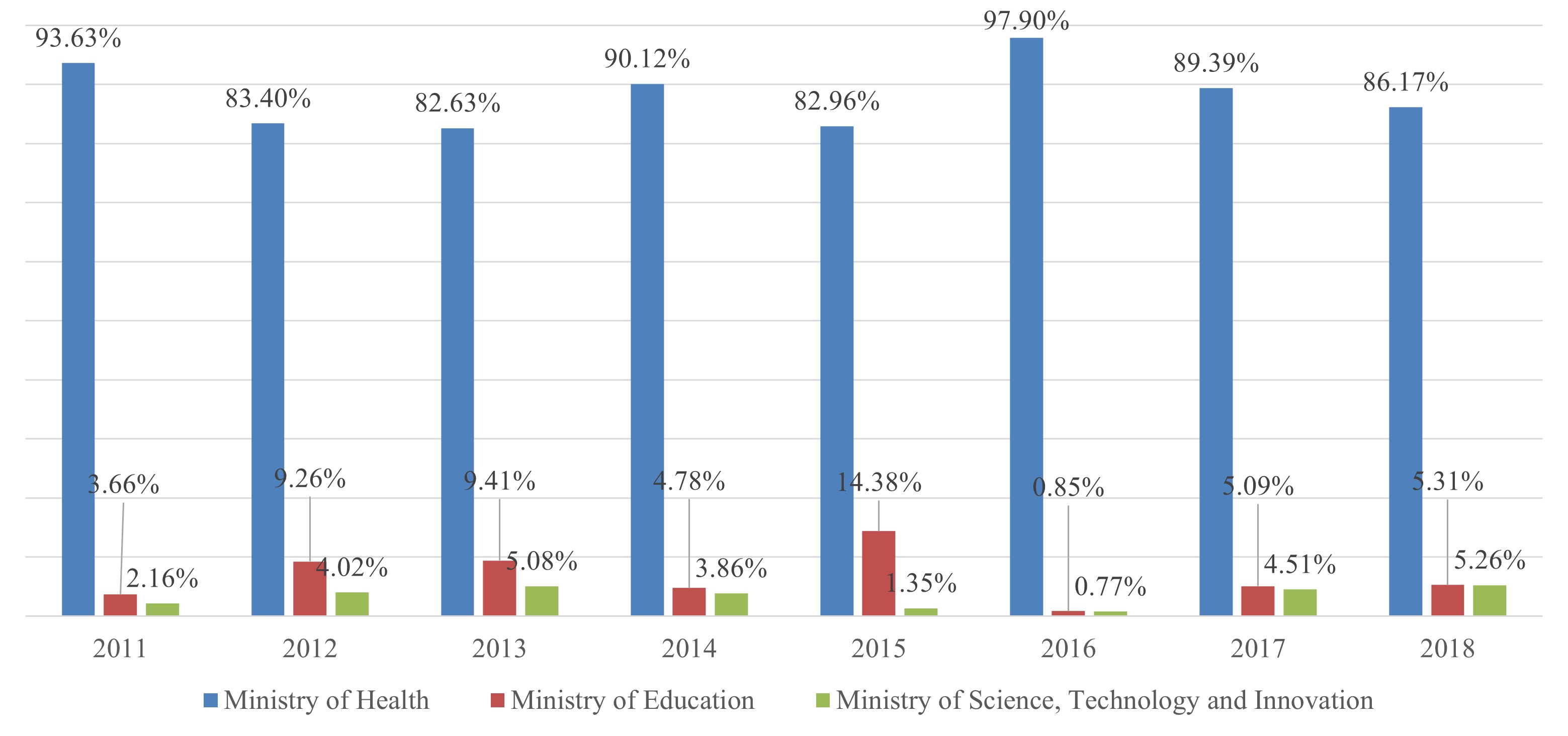}
  \caption{Resources confirmed (\%) to foreign suppliers in the MS, ME and MCTI | Source: Data taken from the Federal Government Procurement Panel}
  \label{fig:fig5}
\end{figure}

The Ministry of Health (MS) was the body that confirmed most resources in favour of foreign suppliers in the period considered, followed by the Ministry of Education (ME) and the Ministry of Science, Technology and Innovation (MCTI). Figure \ref{fig:fig5} compares these three bodies' percentage terms.

\begin{figure}[h]
  \centering
  \includegraphics[width=0.75\textwidth]{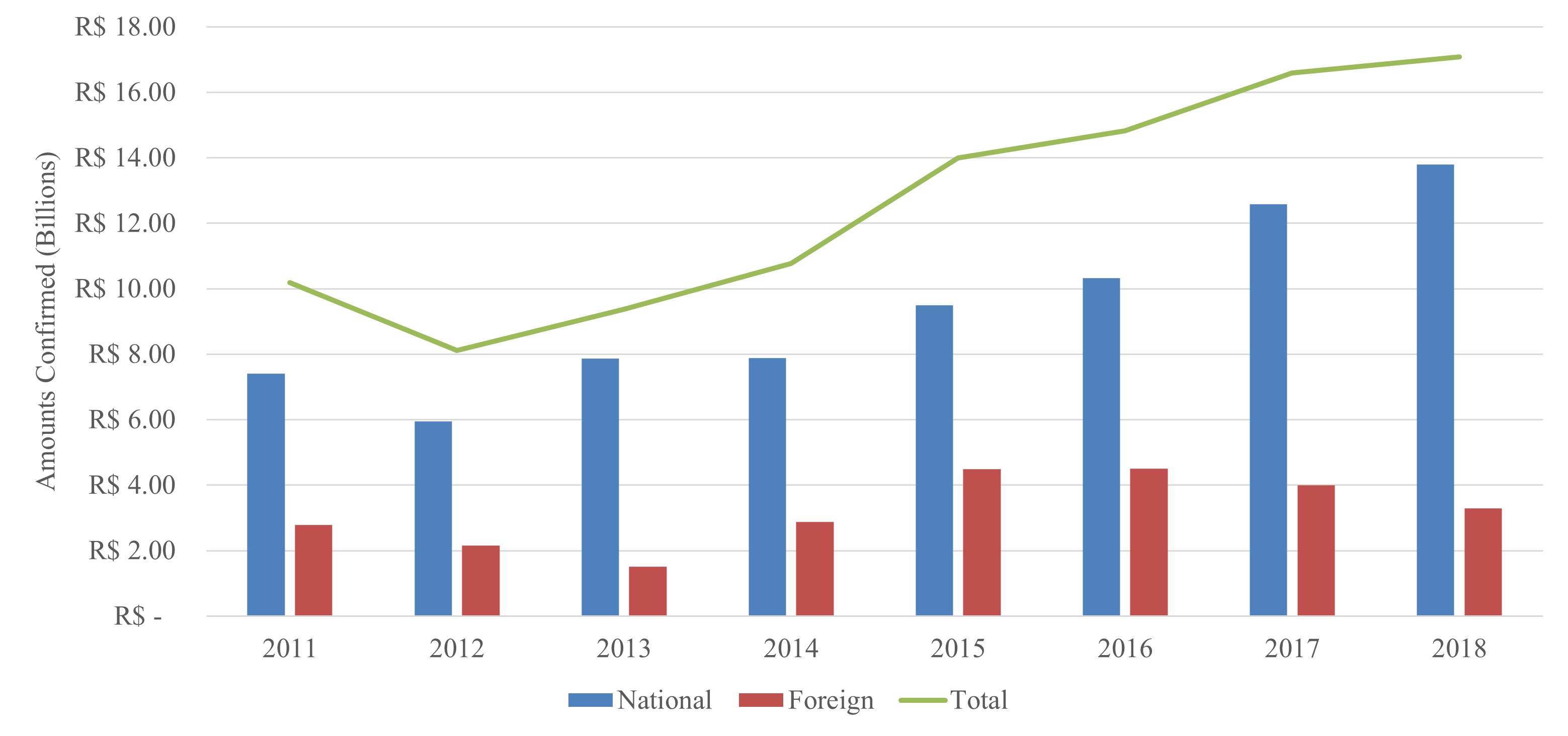}
  \caption{Comparison between resources confirmed to national and foreign suppliers and the total for the MS | Source: Data taken from the Federal Government Procurement Panel}
  \label{fig:fig6}
\end{figure}

Concerning the MH, the total came to about R\$ 25.6 billion, representing approximately 88.67\% of the total resources confirmed for foreigners. Of this total, approximately R\$ 16.3 billion (63.85\%) were confirmed by Impossible Bidding. Figure \ref{fig:fig6} compares the resources confirmed to national and foreign suppliers and the total for this ministry.

\section{Discussion}

According to the OECD, in Brazil, conservative estimates suggest that public procurement accounts for approximately 8.7\% of GDP, where 1.6\% would be attributed to the federal government, 1.5\% to state governments, 2.1\% to municipal governments and 3.2\% to public enterprises and mixed-economy companies (OECD, 2012). However, as we can see in Figure \ref{fig:fig2}, throughout the period, the percentage of public procurement in the federal government did not exceed 1.43\%, below, therefore, the organisation's estimates, which should be viewed with caution since it may be an indication that the federal government may be performing acquisitions below the real needs, which can contribute to the inefficiency and ineffectiveness of public administration.

Two distinct aspects deserve to be analysed concerning how the hiring process was carried out. The first refers to the evolution of confirmed resources in each contracting form. The second refers to the difference between the amounts confirmed through direct hiring (Waivers and Impossible Biddings) and competitive bidding, i.e., using the Reverse Auction (open competitiveness).

Concerning the evolution of confirmed resources, Figure \ref{fig:fig4} shows different behaviours depending on the form of contracting. Regarding Waiver Bidding, there is an upward trend in the amount of resources confirmed to foreigners between 2011 and 2017, with a drop only in 2018 and, about the Reverse Auction, there is a downward trend throughout the period.

\begin{figure}[h]
  \centering
  \includegraphics[width=0.75\textwidth]{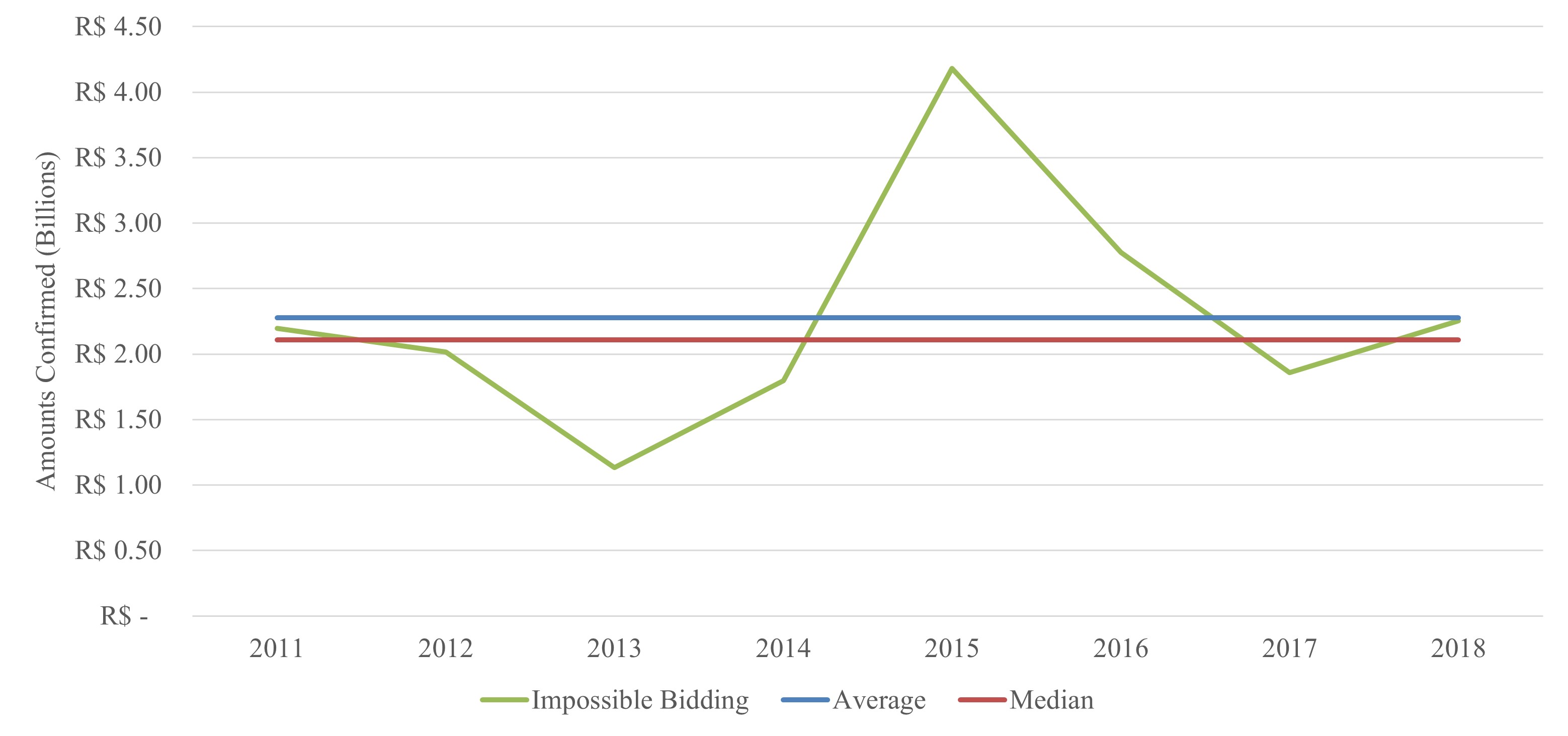}
  \caption{Comparison between resources confirmed to national and foreign suppliers and the total for the MS | Source: Data taken from the Federal Government Procurement Panel}
  \label{fig:fig7}
\end{figure}

The Impossible Bidding has no definite fall or increase trend, as shown in Figure \ref{fig:fig7}. However, there was a significant decrease in resources confirmed for foreign suppliers in 2013 and a significant increase in these resources in 2015 and 2016. It is also observed that the total of approvals to foreigners in the other years remained close to the average and median in the period. At this point, a deeper analysis within the Procurement Panel would allow us to identify and better analyse the nature of these outliers and better understand this variation.

Concerning the second aspect, approvals to foreign suppliers through Impossible Bidding, Waiver Bidding and open competitiveness totalled 99.17\%. In terms of resources, this corresponded to approximately R\$ 18.21 billion (63.07\%) through Impossible Bidding, R\$ 8.73 billion (30.25\%) through Waiver Bidding and R\$ 1.68 billion (5.85\%) through opened competitions, a difference of more than 1,500\% between direct contracting (Waiver and Impossible Bidding) and through opened competitions.

Considering this predominance of direct contracting and also the tendency mentioned above to increase the use of the Waiver of Bidding and decrease the use of the open procedure, it may be of great importance to conduct additional studies on the reasons behind this choice, as well as on the nature of the objects contracted, to understand the real context and, if necessary, propose changes to stimulate competitiveness and allow for increased savings.

It is worth noting that there is a provision for the procedure of registration of foreign suppliers in the Unified Suppliers Registration System (SICAF), which would allow - if competition is feasible - these companies to participate in quotations and procurement procedures in electronic form (Brazil, 2019f, c). Moreover, if the competition is viable but not feasible to use the electronic form of Reverse Auction, it could justifiably be opted for face-to-face, as established in paragraph 4 of the art. 1 of Federal Decree N.º 10,024/2019. In addition to the competitiveness among possible participants, this would bring direct negotiation with the auctioneer in the manner established by item XVI of art. 11 of Federal Decree N.º 3,555/2000, which would undoubtedly increase the possibility of price reduction.

The Ministry of Health alone accounted for about R\$ 100.9 billion (23.88\%) of the total resources confirmed and about R\$ 25.6 billion (88.67\%) confirmed for foreigners. Figure \ref{fig:fig6} compares total and confirmed foreign and domestic resources yearly. In this sense, two distinct scenarios call our attention. The first scenario refers to the downward trend, since 2015, in the volume of resources confirmed to foreign suppliers and about which there is insufficient information to assess its causes.

The second scenario, which is more complex, concerns the total resources confirmed, which have shown a tendency to increase since 2012, with an average of 13.53\% per year, in line with the growth in health spending that occurs worldwide.

According to OECD data, in 2016, global health spending reached US\$ 7.5 trillion, representing close to 10\% of global GDP, with the percentage of GDP ranging, on average, between 6.3\% (lower-middle-income countries) and 8.2\% (high-income countries). Global health spending has increased every year, growing, in real terms, at an average annual rate of 4.0\%, faster than the growth of the worldwide economy, which was 2.8\% per year over the same period. Moreover, health spending grew more quickly in lower-middle-income countries, with, on average, about 6.0\% or more annually (Xu; Soucat; Kutzin, 2018).

This increase can perhaps be explained by the set of some factors such as a diverse group of agents, a dynamic internal and external environment (Moons; Waeyenbergh; Pintelon, 2019), product complexity (Abdulsalam et al., 2015; Moons; Waeyenbergh; Pintelon, 2019), the fact that physicians play an essential role in decisions regarding drug purchasing (Schneller; Smeltzer, 2006), product diversity and criticality, and the very mission of healthcare organisations (Abdulsalam et al., 2015), which make healthcare systems possess unique characteristics. They are complex systems that require an adequate flow of products and services to satisfy the needs of those who serve patients (Schneller; Smeltzer, 2006), and in search of satisfaction of these needs, it will be increasingly necessary to develop strategies that reduce costs and, at the same time, make the service to users more efficient and effective.

More specific studies could help understand the nature of this movement, which may have several origins, such as, for example, the import process being carried out by national representatives with a guarantee of exclusive sale in the domestic market (probability of increasing the price charged to the government) or the development of products in the domestic market with equivalent standards to those imported and therefore able to meet the needs of the public health system (probability of decreasing the price charged). These studies could ultimately lead to inhibiting or incentive mechanisms to safeguard the public interest.

\section{Conclusion}

The main objective of this article was to discuss the weight of the participation of foreign suppliers in the context of federal bids between 2011 and 2018. The results obtained through research with the Procurement Panel and the discussion and ideas generated from these results allowed us to conclude this objective.

Thus, the Procurement Panel proved to be a helpful tool, easy to access and use and that meets the role it proposes, which is to present in one place the central figures of public procurement and offer an overview of public spending and bidding behaviour within the federal administration. However, there are still some limitations in terms of information, which will be discussed below.

As for the weight of foreign participation, although, in percentage terms, only 6.83\% of the total resources between 2011 and 2018 were confirmed to these suppliers, in financial terms, this percentage corresponds to about R\$ 28.9 billion. This is a relevant sum and shows how attractive the Brazilian public market can be for foreign companies to sell their products. To get a better idea of the importance of this amount, according to IBGE data (Brazil, 2019e), 4,868 municipalities, out of a total of 5,099, had accumulated GDP between 2010 and 2017, lower than the amount confirmed to foreign suppliers in federal processes between 2011 and 2018.

In this sense, further studies on several fronts besides those mentioned throughout the text should be conducted to understand more clearly the mechanisms that may encourage and/or restrict the participation of foreign suppliers in Brazilian biddings, given that, in many areas, such as health, foreign companies can help in the process of incorporating more modern technologies, without harm to the local market, since they are not available domestically.

Finally, it is expected that this article will fill a gap in quantitative studies on international bidding and serve as an incentive for new publications on the subject. At the same time, it is expected that this article will contribute not only to the scientific community but mainly to public managers so that they can generate improvements in increasing competition based on the results described. The results show no competition in most bids, which can hinder studies around price analysis and determination of prices/reference values. In this sense, the increased use of competitive bidding modalities, especially the Reverse Auction, could mitigate overbilling and/or overpricing and generate an increase in efficiency and effectiveness in the expenditure of public resources.

\section{Limitations of this study}

Since the Procurement Panel was developed to offer information from all agencies that make up SISG (Brazil, 2019b) and that the agencies and entities of the direct, autarchic and foundational federal administration are part of SISG, as provided in §1, of art. 1, of Federal Decree N.º 1,094/94 (Brazil, 1994), in principle, acquisitions made by state-owned companies (public companies and mixed-economy companies) are not included in these resources. Thus, since Brazil had 130 state-owned companies in 2018 (Brazil, 2018), the participation of foreign suppliers in Brazilian bids may vary (for more or less), a fact that deserves a more particular and in-depth study.

In addition, there is no information available in the Procurement Panel to identify whether the acquisitions were made with resources from international bodies of which Brazil is part (§ 5º of art. 42 of the Bidding Law) or from an international agreement (art. 24, item XIV, of the Bidding Law), nor to verify whether, for these hypotheses, there was the participation of foreign suppliers, so improvements in this direction could be implemented to facilitate future analyses.

\section*{References}

Abdulsalam, Y. et al. Health care matters: supply chains in and of the health sector, Journal of Business Logistics, v. 36, n. 4, p. 335–339, 2015. DOI: 10.1111/jbl.12111.

Bittencourt, S. Reverse Auction eletrônico: Decreto no 5.450, de 31 de maio de 2005, Lei no 10.520, de 17 de julho de 2002, considerando também a Lei Complementar no 123/2006, que estabelece tratamento diferenciado e favorecido às microempresas e empresas de pequeno porte. 3ª ed. Rev ed. Belo Horizonte, Fórum, 2010.

Bittencourt, S. Reverse Auction presencial: comentários ao Decreto n.o 3.555/2000 e ao regulamento do Reverse Auction, atualizado pelo Decreto n.o 7.174/2010: considerando as Leis n.os 10.520/2002 e 8.666/1993 atualizadas. 2ª ed. rev ed. Belo Horizonte, Fórum, 2012.

Brasil. Decreto n.º 1.094, de 23 de março de 1994 - dispõe sobre o Sistema de Serviços Gerais (SISG) dos órgãos civis da administração federal direta, das autarquias federais e fundações públicas, e dá outras providências. 1994. Presidência da República. Available in: \href{http://www.planalto.gov.br/ccivil_03/decreto/Antigos/D1094.htm}{link}. Accessed in: 28 dez. 2019. 

Decreto n.º 10.024, de 20 de setembro de 2019 - regulamenta a licitação, na modalidade Reverse Auction, na forma eletrônica, para a aquisição de bens e a contratação de serviços comuns, incluídos os serviços comuns de engenharia, e dispõe sobre o uso da dispensa el. 2019a. Presidência da República. Available in: \href{http://www.planalto.gov.br/ccivil_03/_ato2019-2022/2019/decreto/D10024.htm}{link}. Accessed in: 28 dez. 2019.

Decreto n.º 5.450 de 31 de maio de 2005 - regulamenta o Reverse Auction, na forma eletrônica, para aquisição de bens e serviços comuns, e dá outras providências. 2005a. Presidência da República. Available in: \href{http://www.planalto.gov.br/ccivil_03/_ato2004-2006/2005/decreto/d5450.htm}{link}. Accessed in: 28 dez. 2019.

Decreto n.º 5.504 de 05 de agosto de 2005 - estabelece a exigência de utilização do Reverse Auction, preferencialmente na forma eletrônica, para entes públicos ou privados, nas contratações de bens e serviços comuns, realizadas em decorrência de transferências vol. 2005b. Presidência da República. Available in:  \href{http://www.planalto.gov.br/Ccivil_03/_ato2004-2006/2005/Decreto/D5504.htm}{link}. Accessed in: 28 dez. 2019.

Lei n.º 12.349 de 15 de dezembro de 2010 - altera as Leis nºs 8.666, de 21 de junho de 1993, 8.958, de 20 de dezembro de 1994, e 10.973, de 2 de dezembro de 2004; e revoga o § 1º do art. 2º da Lei nº 11.273, de 6 de fevereiro de 2006. 2010. Presidência da República. Available in: \href{http://www.planalto.gov.br/ccivil_03/_ato2007-2010/2010/lei/l12349.htm}{link}. Accessed in: 28 dez. 2019.

Lei n.º 8.666 de 21 de junho de 1993 - regulamenta o art. 37, inciso XXI, da Constituição Federal, institui normas para licitações e contratos da administração pública e dá outras providências. 1993. Presidência da República. Available in: \href{http://www.planalto.gov.br/ccivil_03/leis/L8666compilado.htm}{link}. Accessed in: 28 dez. 2019.

Painel de Compras. 2019b. Ministério da Economia, Planejamento, Desenvolvimento e Gestão. Available in: \href{http://paineldecompras.planejamento.gov.br/QvAJAXZfc/opendoc.htm?document=paineldecompras.qvw&lang=en-US&host=QVS%40srvbsaiasprd04&anonymous=true.}{link}. Accessed in: 28 dez. 2019.

Reverse Auction Eletrônico - fornecedor. 2019c. Ministério da Economia, Planejamento, Desenvolvimento e Gestão. Available in: \href{https://www.comprasgovernamentais.gov.br/index.php/pregaoeletronico-fornecedor-faq#P21}{link}. Accessed in: 28 dez. 2019.

Produto Interno Bruto - PIB. 2019d. Instituto Brasileiro de Geografia e Estatística - IBGE. Available in: \href{https://www.ibge.gov.br/explica/pib.php}{link}. Accessed in: 28 dez. 2019.

Produto Interno Bruto dos Municípios. 2019e. Instituto Brasileiro de Geografia e Estatística - IBGE. Available in: \href{https://www.ibge.gov.br/estatisticas/economicas/9088-produto-internobruto-dos-municipios.html?=&t=resultados}{link}. Accessed in: 28 dez. 2019.

Relatórios Empresas Estatais Federais. 2018. Ministério da Economia, Planejamento, Desenvolvimento e Gestão. Available in: \href{http://www.planejamento.gov.br/assuntos/empresasestatais/191212_EmpresasEstataisFederaisCNPJsedeedatasdecriaoeconstituioInformaesSIEST.pdf}. Accessed in: 28 dez. 2019.

SICAF - 100\% Digital. 2019f. Ministério da Economia, Planejamento, Desenvolvimento e Gestão. Available in: \href{https://www.comprasgovernamentais.gov.br/index.php/sicaf-100digitalfaq#AS14}{link}. Accessed in: 28 dez. 2019.

SISG. 2019g. Ministério da Economia, Planejamento, Desenvolvimento e Gestão. Available in: https://www.comprasgovernamentais.gov.br/index.php/sisg. Accessed in: 28 dez. 2019.

Carvalho, E. de. Licitações Internacionais no Direito Brasileiro [S.l.], Jusbrasil. Available in: \href{https://eleazaralbuquerquedecarvalho.jusbrasil.com.br/artigos/154576563/licitacoesinternacionais-no-direito-brasileiro}{link}. 2014.

Castro, H. B. de. As Licitações Internacionais no Ordenamento Jurídico Brasileiro, com ênfase no Princípio da Competitividade. 2002. 191 f. Universidade Católica de Brasília - UCB, 2002. Available in: \href{https://bdtd.ucb.br:8443/jspui/handle/123456789/416}{link}.

Faria, E. R. de et al. “Fatores determinantes na variação dos preços dos produtos contratados por Reverse Auction eletrônico”, Revista de Administração Pública, v. 44, n. 6, p. 1405–1428, dez. 2010. DOI:10.1590/S0034-76122010000600007.

Filho, M. J. Reverse Auction. Comentários A Legislação do Reverse Auction comum e eletrônico. 6ª ed. rev ed. São Paulo, Dialética, 2013.

Inamine, R.; Erdmann, R. H.; Marchi, J. J. “Análise do sistema eletrônico de compras do governo federal brasileiro sob a perspectiva da criação de valor público”, Revista de Administração, v.47, n. 01, p. 129–139, 2012. DOI: 10.5700/rausp1030.

Marrara, T.; Campos, C. S. “Licitações internacionais: regime jurídico e óbices à abertura do mercado público brasileiro a empresas estrangeiras”, Revista de Direito Administrativo, v. 275, p.155, 29 ago. 2017. DOI: 10.12660/rda.v275.2017.71651.

Moons, K.; Waeyenbergh, G.; Pintelon, L. “Measuring the logistics performance of internal hospital supply chains – a literature study”, Omega, v. 82, p. 205–217, jan. 2019. DOI: 10.1016/j.omega.2018.01.007.

Moreira, E. B.; Guimarães, B. S.; Torgal, L. “Licitação internacional e empresa estrangeira: os cenários brasileiro e europeu”, Revista de Direito Administrativo, v. 269, p. 67, 11 nov. 2015. DOI:10.12660/rda.v269.2015.57595.

Oecd. OECD Integrity Review of Brazil. 1ª ed. Paris, OECD Publishing, 2012. Available in: \href{https://www.oecd-ilibrary.org/governance/brazil-oecd-integrity-review_9789264119321-en}{link}. (OECD Public Governance Reviews).

Oliveira, R. C. R. Licitações e Contratos Administrativos. Teoria e Prática. 5ª ed. rev ed. Rio de Janeiro; São Paulo, Forense; Método, 2015.

Peixoto. É possível fazer Reverse Auction eletrônico internacional? Portal licitacao.com.br. Available in: \href{http://licitacao.uol.br/adm/img_upload/asse237.pdf}{link}. Accessed in: 13 nov. 2009.

Pereira, L. H. de C. Licitações internacionais e a Lei no 8.666/93. 1ª ed. São Paulo, All Print Editora, 2013.

Pestana, M. Licitações públicas no Brasil: exame integrado das Leis 8.666/1993 e 10.520/2002. 1ª ed. São Paulo, Atlas, 2013.

Pighini, B. C.; Gomes, M. F. “Da licitação internacional e da margem de preferência”, Revista FMU Direito, v. 40, p. 38–54, 2013. Available in: \href{http://www.revistaseletronicas.fmu.br/index.php/FMUD/article/view/429/596}{link}.

Ribeiro, G. L. V. A evolução da Licitação. Revista Contábil e Empresarial Fiscolegis. Maio 2007. Available in: \href{http://www.netlegis.com.br/indexRC.jsp?arquivo=detalhesArtigosPublicados.jsp&cod2=854}{link}. Accessed in: 12 mar. 2010.

Ribeiro, L. C.; Pereira, D. S. “Direito administrativo global, financiamentos internacionais e licitações públicas”, Revista de Contratos Públicos – RCP, v. 8, p. 111–132, 2016. Available in: \href{https://s3.amazonaws.com/academia.edu.documents/50971006/Direito_administrativo_global__financiamentos_internacionais_e_licitacoes_publicas.pdf?response-contentdisposition=inline%3B filename%3DDireito_administrativo_global_financiame.pdf&X-Amz-Algorithm=.}{link}

Schneller, E. S.; Smeltzer, L. R. Strategic Management of the Health Care Supply Chain. First ed. San Francisco, CA, John Wiley \& Sons, Ltd, 2006.

Schwind, R. W. Licitações internacionais: participação de estrangeiros e licitações realizadas com financiamento externo. 1ª ed. Belo Horizonte, Fórum, 2013.

Silveira, E. S. et al. “Análise do processo de compras do setor público: o caso da Prefeitura Municipal de Dourados/MS”, Revista de Administração IMED, v. 2, n. 3, p. 158–171, 30 dez. 2012. DOI: 10.18256/2237-7956/raimed.v2n3p158-171.

Teixeira, J. C.; Penedo, A. S. T.; Almeida, R. de. “A evolução do processo licitatório com ênfase nos conceitos de economia versus qualidade”, Nucleus, v. 9, n. 2, p. 335–349, 31 out. 2012. DOI: 10.3738/1982.2278.586.

Xu, K; Soucat, A.; Kutzin, J. et al. Public spending on health: a closer look at global trends. Geneva, World Health Organization. 2018.

\end{document}